\documentclass{PoS}

\rightline{\sffamily RBRC-615}\vspace*{-10mm}

\title{
Calculation of $\Delta I = 3/2$ kaon weak matrix elements including two-pion interaction effects in finite volume}

\ShortTitle{Calculation of $\Delta I = 3/2$ kaon weak matrix elements ....}

\author{\speaker{Takeshi Yamazaki}
        \thanks{Present address: University of Connecticut, Physics Department,
                U-3046 2152 Hillside Road, Storrs, Connecticut 06269-3046, US}
        \ for RIKEN-BNL-Columbia Collaboration\\
        RIKEN BNL Research Center, 
        Brookhaven National Laboratory\\
        Upton, New York 11973, US\\
        E-mail: \email{yamazaki@quark.phy.bnl.gov}}

\abstract{
We calculate $\Delta I = 3/2$ kaon decay matrix elements using domain 
wall fermions and the DBW2 gauge action at one coarse lattice spacing 
corresponding to $a^{-1} = 1.3$ GeV. 
We employ the Lellouch and L\"uscher formula and its extention 
for non-zero total momentum to extract the infinite volume, 
center-of-mass frame decay amplitudes.  
The decay amplitudes obtained from the methods correspond to
those from the indirect method with full order chiral perturbation theory.
We confirm that the result is consistent with the previous result
calculated with H-parity (anti-periodic) boundary condition 
by investigating the relative momentum dependence.
We evaluate the decay amplitude Re$A_2$
at the physical point by a chiral extrapolation with 
a polynomial function of $m_\pi^2$ and the relative momentum
as well as the $\Delta I = 3/2$ electroweak penguin contributions 
to $\varepsilon^\prime / \varepsilon$.
We found that the result of Re$A_2$ reasonably agrees with the experiment.
}

\FullConference{XXIV International Symposium on Lattice Field Theory\\
		 July 23-28 2006\\
		 Tucson Arizona, US}

\begin{document}

\section{Introduction}

It is difficult to directly calculate 
the $K \to \pi\pi$ weak matrix elements on lattice
due to difficulties of calculation of the two-pion state in finite volume.
Indirect method~\cite{indirect_method} is one of the candidates
to avoid the difficulty.
In the indirect method $K\to\pi\pi$ process is reduced to $K\to\pi$ and 
$K\to 0$ processes through chiral perturbation theory (ChPT).
The RBC~\cite{RBC} and CP-PACS~\cite{CP-PACS} Collaborations calculated 
full non-leptonic kaon decay processes with the method. 
Their final results of $\varepsilon^\prime / \varepsilon$, however,
have the opposite sign to the experiment.
In the calculations there are many systematic errors, {\it e.g.},
calculating at one finite lattice spacing with quenched approximation,
and using the reduction with tree level ChPT.
The indirect method might cause larger systematic errors than other sources,
because the final state interaction of the two-pion
is expected to play an important role in the decay process.
Thus we have to treat the scattering effect of the two-pion state
directly on lattice to eliminate this systematic error.

There are two main difficulties for the direct method,
where the two-pion state is calculated on lattice.
The one is to extract the two-pion state contribution
with non-zero relative momentum in the $K\to\pi\pi$
four-point function, which was pointed out by Maiani and Testa~\cite{MT}.
So far there are several ideas for solving the problem.
A method is employed with a proper projection of the $K\to\pi\pi$
four-point functions~\cite{NI}.
In the method we need complicated calculations and analyses, 
{\it e.g.}, diagonalization of a matrix of 
the two-pion correlation functions~\cite{LW},
to treat the two-pion state with non-zero relative momentum on lattice.
A simpler idea, where complicated analyses are not required,
is to prohibit the zero momentum two-pion state.
Recently Kim~\cite{CK} reported an exploratory study 
of the idea with H-parity (anti-periodic) boundary conditions
in the spatial direction.
He succeeded to extract the two-pion state with non-zero relative momentum
from the ground state contribution of correlation functions,
because the zero momentum two-pion state is prohibited by 
the boundary condition.
We can also forbid the zero momentum two-pion state 
in center-of-mass (CM) frame
by performing the calculation 
in non-zero total momentum (Lab) frame, $|\vec{P}|\ne 0$.
In the frame the ground state of the two-pion is $|\pi(0)\pi(\vec{P})\rangle$,
which is related to the two-pion state with the non-zero relative momentum
in the CM frame.
Thus, we can extract the two-pion state with non-zero momentum from
the ground state contributions~\cite{BGLLMPS} as well as
in the H-parity boundary case.

The other difficulty is related to the finite volume correction due to
the two-pion interaction.
We have to pay attention to the finite volume effect of the two-pion to 
obtain matrix elements in the infinite volume,
because it is much larger than that of a one-particle state.
Lellouch and L\"uscher (LL)~\cite{LL} suggested a solution which is a relation 
between the finite and infinite volume, center-of-mass frame decay amplitudes.
However, this relation is valid only in the CM frame with
periodic boundary condition in the spatial direction,
so that we need a modified formula when we utilize H-parity
boundary condition~\cite{CK} or Lab frame calculation.
Recently, two groups, Kim {\it et al.}~\cite{KSS} and 
Christ {\it et al.}~\cite{CKY}, suggested a formula which is an extension 
of the LL formula for the Lab frame calculation.

Here we attempt to apply these two methods, Lab frame calculation and the 
extended formula, to the calculation of the $\Delta I = 3/2$ kaon weak 
matrix elements.
Our preliminary result was presented in Ref.~\cite{LAT05}.
The results presented here are systematically larger than the
preliminary results reported in last year lattice conference.  During
the past year we discovered that larger time separations were needed
to remove excited state contamination.

\section{Methods}

Through the extended formula~\cite{KSS,CKY},
the infinite volume decay amplitude $|A|$ in the CM frame
is given by
\begin{equation}
| A |^2 = 8\pi \gamma^2 \left(\frac{E_{\pi\pi}}{p}\right)^3
\left\{ p \frac{ \partial \delta(p) }{ \partial p }
      + q \frac{ \partial \phi_{\vec{P}}(q) }{ \partial q }
\right\} | M |^2,
\label{eq_Lab_Formula}
\end{equation}
where $|M|$ is the finite volume, Lab frame decay amplitude,
$\gamma$ is a boost factor, 
$E_{\pi\pi}$ is the CM two-pion energy,
and $\delta$ is the scattering phase shift
of the final state interaction.
The relative momentum $p^2$ is defined by 
the two-pion energy, $p^2 = E_{\pi\pi}^2/4 - m_\pi^2$.
The function $\phi_{\vec{P}}$ with $\vec{P}$ being the total momentum, 
derived by Rummukainen and Gottlieb~\cite{RG},
is defined by
\begin{equation}
\tan \phi_{\vec{P}}(q) = 
- \frac{\gamma q\pi^{3/2}}{Z^{\vec{P}}_{00}(1;q^2;\gamma)},
\end{equation}
where $q^2 = ( p L / 2\pi )^2$, and
\begin{equation}
Z_{00}^{\vec{P}}(1;q^2;\gamma) = 
\frac{1}{\sqrt{4\pi}}
\sum_{\vec{n}\in \mathbb{Z}^3}\frac{1}{n_1^2+n_2^2+\gamma^{-2}(n_3+1/2)^2-q^2},
\label{eq_phi}
\end{equation}
in the $\vec{P} = (0, 0, 2\pi/L)$ case.
The formula eq.~(\ref{eq_Lab_Formula})
is valid only for on-shell decay amplitude,
{\it i.e.}, $E_{\pi\pi} = m_K$ as in LL formula~\cite{LL}.

We calculate the four-point function for $\Delta I = 3/2$ $K\to\pi\pi$ decay
with total momentum $\vec{P} = \vec{0}$ and $(0, 0, 2\pi/L)$.
The four-point function $G_i(\vec{P};t,t_\pi,t_K)$ is defined by
\begin{equation}
G_i(\vec{P};t,t_\pi,t_K) = 
\langle 0 | 
[ K^0(\vec{P};t_K) ]^\dagger O^{3/2}_i(t) 
\pi^+\pi^-(\vec{P};t_\pi)
| 0 \rangle,
\end{equation}
where the operators $O^{3/2}_i$ are lattice operators entering
$\Delta I = 3/2$ weak decays
\begin{eqnarray}
O^{3/2}_{27,88} &=& 
(\overline{s}^ad^a)_L
\left[(\overline{u}^bu^b)_{L,R}-(\overline{d}^bd^b)_{L,R}\right]
       + (\overline{s}^au^a)_L(\overline{u}^bd^b)_{L,R}\ \ ,\\
O^{3/2}_{m88} &=& 
(\overline{s}^ad^b)_L
\left[(\overline{u}^bu^a)_{R}-(\overline{d}^bd^a)_{R}\right]
       + (\overline{s}^au^b)_L(\overline{u}^bd^a)_{R}\ \ ,
\end{eqnarray}
with $(\overline{q}q)_L = \overline{q}\gamma_\mu(1-\gamma_5)q$,
$(\overline{q}q)_R = \overline{q}\gamma_\mu(1+\gamma_5)q$,
and $a,b$ being color indices.
$O^{3/2}_{27}$ and $O^{3/2}_{88}$
are the operators in the (27,1) and (8,8) representations of 
$SU(3)_L\otimes SU(3)_R$ with $I=3/2$, respectively.
$O^{3/2}_{m88}$ is identical to $O^{3/2}_{88}$, except the color indices
are mixed.
We also calculate the four-point function of two pions 
$G_{\pi\pi}(\vec{P};t,t_\pi)$
and the two-point function of the kaon $G_K(\vec{P};t,t_K)$
with zero and non-zero total momenta,
to obtain the needed energies and amplitudes.

\section{Simulation parameters}

We employ the domain wall fermion action with the domain wall
height $M=1.8$, the fifth dimension length $L_s = 12$
and the DBW2 gauge action with $\beta = 0.87$ corresponding
to $a^{-1} = 1.31(4)$ GeV and $m_{\mathrm{res}} = 1.25(3)\times 10^{-3}$.
The lattice size is $L^3 \cdot T = 16^3 \cdot 32$, 
where the physical spatial size corresponds to about 2.4 fm.
Our simulation is carried out at four $u,d$ quark masses, 
$m_u = 0.015, 0.03, 0.04$ and 0.05,
for the chiral extrapolation of the decay amplitudes,
using 252 gauge configurations, except for the lightest mass
where we use 370 configurations.
For interpolations of the amplitudes to the on-shell point,
we calculate the decay amplitudes with 
six strange quark masses, $m_s = 0.12, 0.18, 0.24, 0.28, 0.35$ and 0.44,
at the heavier three $m_u$, 
while we use the three lighter $m_s$ at the lightest $m_u$.
It is enough for the on-shell interpolation with the three $m_s$
in the lightest $m_u$ case.
We will see the point in a later section.

We fix the two-pion operator at $t_\pi = 0$,
and the kaon operator $t_K = 20$
to avoid contaminations from excited states as much as possible.
A quark propagator is calculated by averaging 
quark propagators with periodic and anti-periodic boundary conditions for 
the time direction to obtain a propagator with $2T$ periodicity.
We calculate the quark propagators with Coulomb gauge-fixed wall and 
momentum sources.

\begin{figure}[!t]
\begin{center}
\scalebox{0.30}[0.27]{
\rotatebox{270}{
\includegraphics{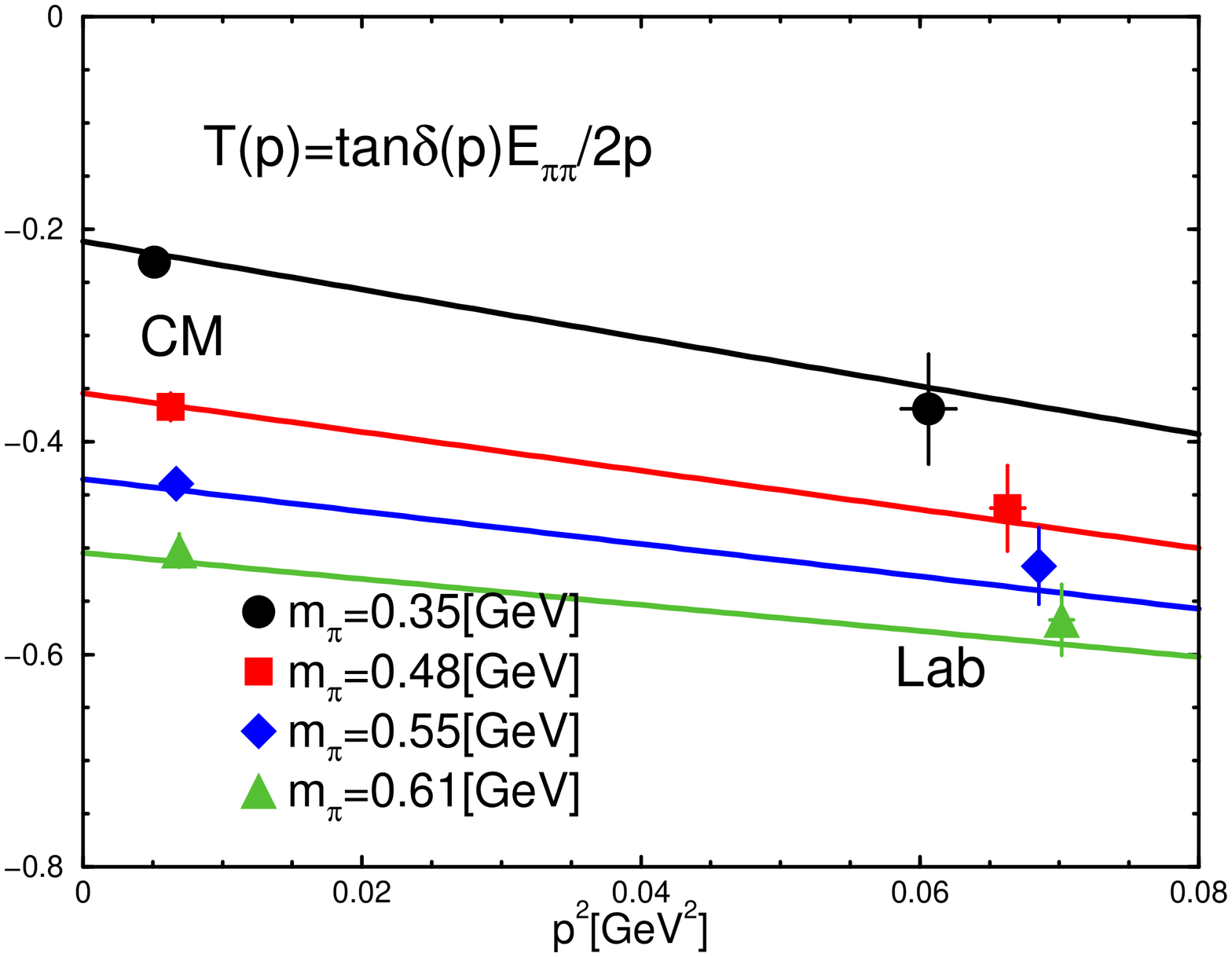}}}
\scalebox{0.30}[0.27]{
\rotatebox{270}{
\includegraphics{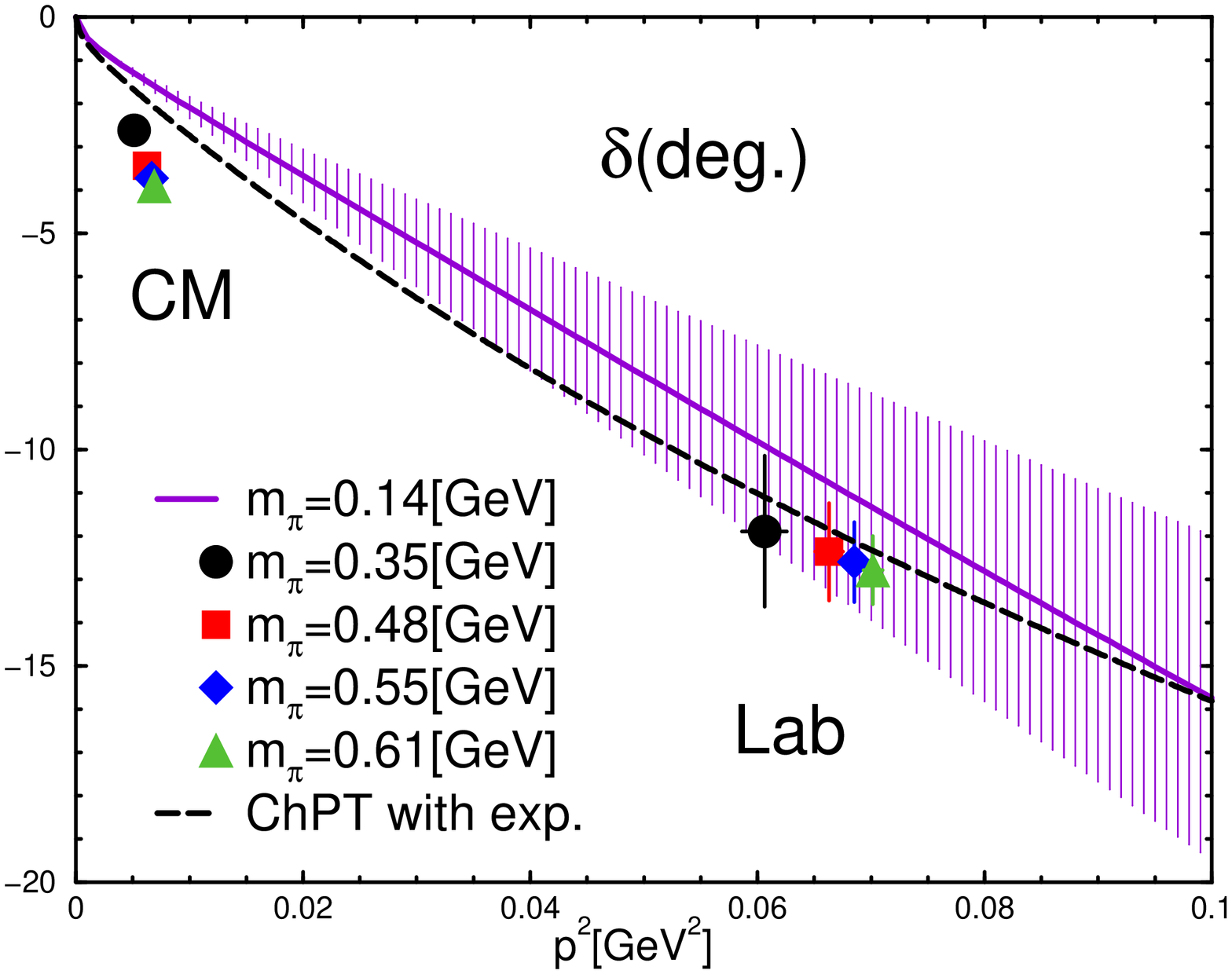}}}
\vspace{-5mm}
\end{center}
\caption{Measured values of scattering amplitude $T$.
and $\delta$ obtained from CM and Lab calculations.
Solid lines are fit results.
\label{fig_delta}}
\end{figure}

\section{Results}

The final state of the $\Delta I = 3/2$ kaon decay is 
the S-wave isospin $I=2$ two-pion state.
The scattering phase shift of the two-pion state can be obtained 
by the finite volume formulae for the CM~\cite{FVF} and
the Lab frames~\cite{RG} with the two-pion energy in each frame.
We define "scattering amplitude''
$T(p) = \tan \delta(p) E_{\pi\pi} / 2 p$,
where $E_{\pi\pi}$ is the two-pion energy in the CM frame.
The scattering amplitude is used for the chiral extrapolation
of the phase shift with
a global fitting for $m_\pi^2$ and $p^2$ with a polynomial function 
$a_{10} m_\pi^2 + a_{20} m_\pi^4 + a_{01} p^2 + a_{11} m_\pi^2 p^2$.
While $p^4$ and $m_\pi^4$ should both be treated as second-order in ChPT, 
for our calculation we have only two different relative momenta.
For that reason we omit the additional $p^4$ term.
The figure~\ref{fig_delta} shows 
the result of $T(p)$ and the fit results at each pion mass
in the left figure,
and the measured values of $\delta(p)$ in the right figure.
The phase shift at the physical pion mass $m_\pi = 0.14$ GeV,
plotted by solid line with the error band in the right graph, 
is comparable with the prediction from ChPT with experiment.
In order to utilize the extension~eq.(\ref{eq_Lab_Formula})
of the LL formula,
we evaluate the derivative of the phase shift
from the fit result, while the derivative of the function $\phi_{\vec{P}}$
is obtained by a numerical derivative.

All the off-shell decay amplitudes are determined by the ratio of
correlation functions with $i = 27, 88$ and $m88$,
$\sqrt{3} G_i(\vec{P};t,t_\pi,t_K) Z_{\pi\pi} Z_K / 
G_{\pi\pi}(\vec{P};t,t_\pi) G_K(\vec{P};t,t_K) $,
where 
$Z_{\pi\pi}$ and $Z_K$ are the overlap of the relevant operator with each state.
We determine the off-shell amplitude in the flat region of the ratio
as a function of $t$,
because the ratio will be a constant for those values of $t$
when these correlation functions are dominated by each ground state.

\begin{figure}[!t]
\begin{center}
\scalebox{0.30}[0.27]{
\rotatebox{270}{
\includegraphics{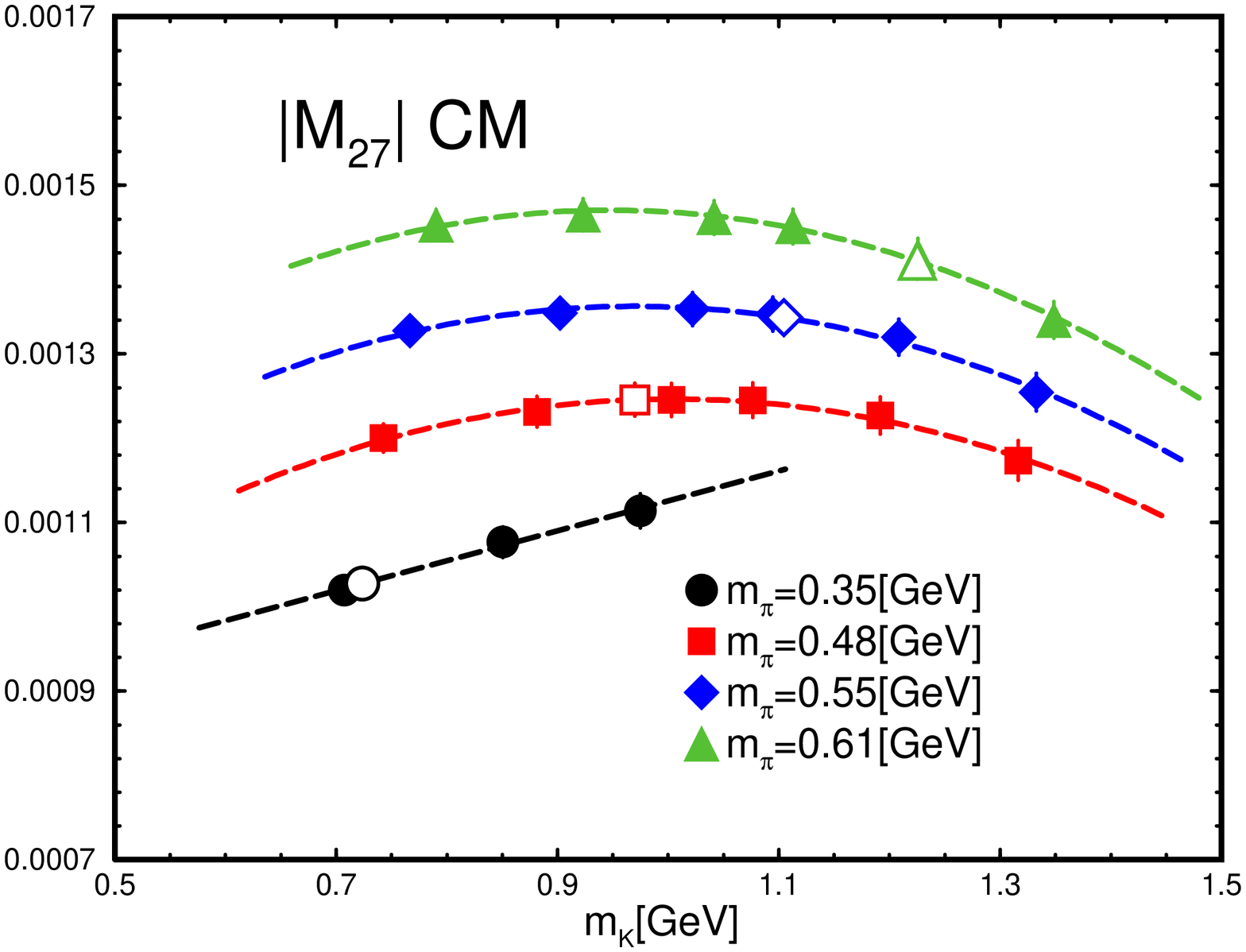}}}
\scalebox{0.30}[0.27]{
\rotatebox{270}{
\includegraphics{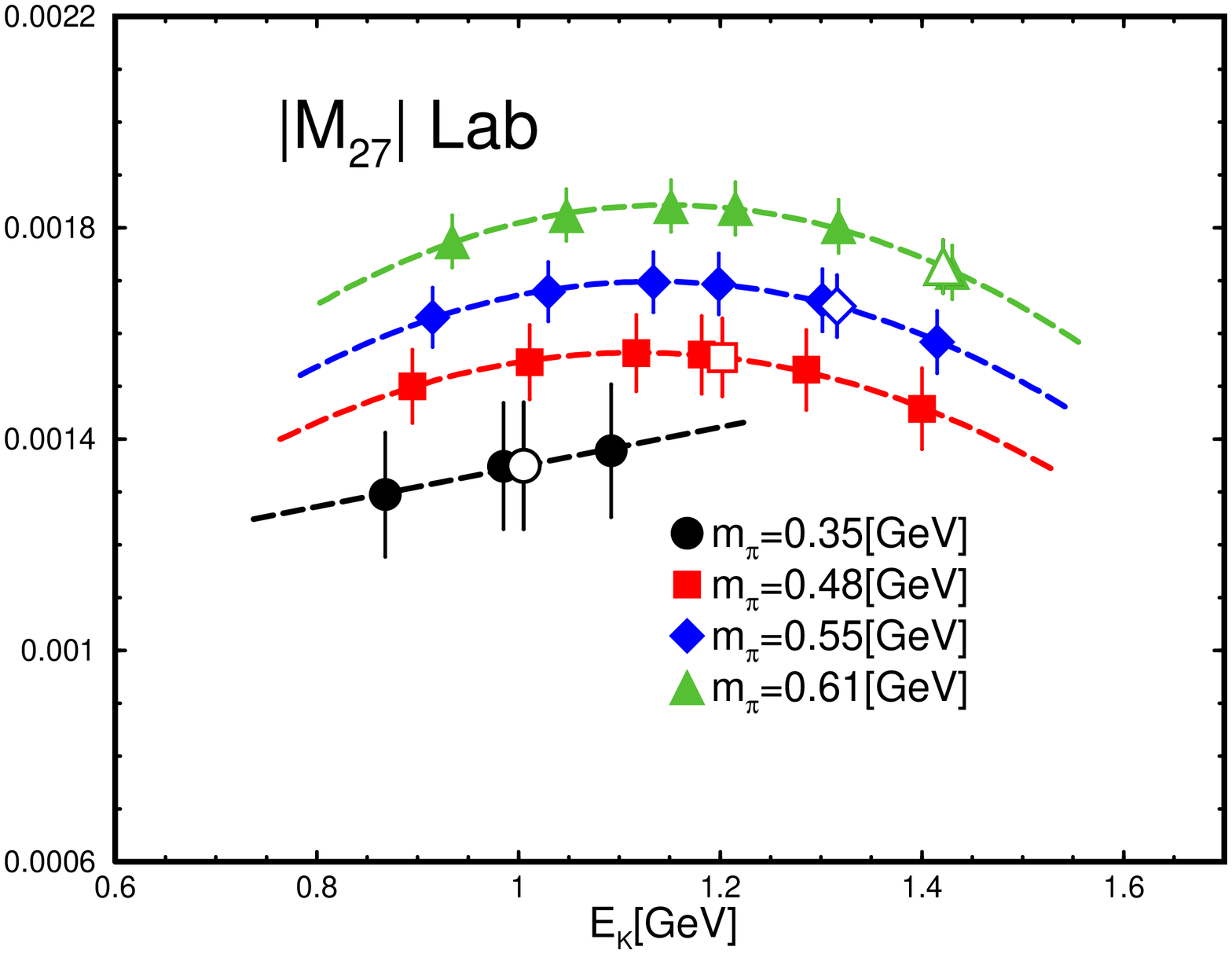}}}
\vspace{-5mm}
\end{center}
\caption{
Interpolation of off-shell decay amplitude to on-shell in CM and Lab
frames. 
Closed(open) symbol is off-shell(on-shell) amplitude.
\label{fig_ext}}
\end{figure}

The figure~\ref{fig_ext} shows the off-shell decay amplitude of 
the $27$ operator in finite volume
and its interpolation to the on-shell $E_K = E_{\pi\pi}$
in the cases of both the frames.
In the lightest pion mass case,
we can linearly interpolate the data with the three kaon energies.
On the other hand, the off-shell decay amplitudes at the heavier pion masses
have large curvature for the kaon energy, so that we fit the data with
a quadratic function.
In all the pion mass cases, the interpolated, on-shell decay amplitude
and the fit line are plotted in the figure.

\begin{figure}[!t]
\begin{center}
\scalebox{0.32}[0.29]{
\rotatebox{270}{
\includegraphics{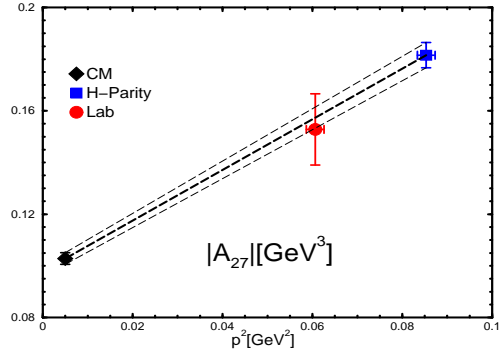}}}
\vspace{-5mm}
\end{center}
\caption{Infinite volume decay amplitude of 27 operator
obtained from different calculation methods.
\label{fig_mom}}
\end{figure}

The decay amplitudes on finite volume in each frame
are converted to those of the CM frame in the infinite volume  
through the LL formula~\cite{LL} 
and its extended formula~eq.(\ref{eq_Lab_Formula})
using the derivatives of $\delta$ and $\phi_{\vec{P}}$ obtained in the above.
In Fig.~\ref{fig_mom}
we plot the infinite volume decay amplitude of the 27 operator
obtained from the different frames at the lightest pion mass
as a function of $p^2$.
The previous result obtained from H-parity boundary calculation
is also plotted to compare with these results.
The amplitude obtained from the Lab frame is consistent with the line
determined from those of the CM frame and H-parity boundary calculations.

%
\begin{figure}[!h]
\begin{center}
\scalebox{0.32}[0.29]{
\rotatebox{270}{
\includegraphics{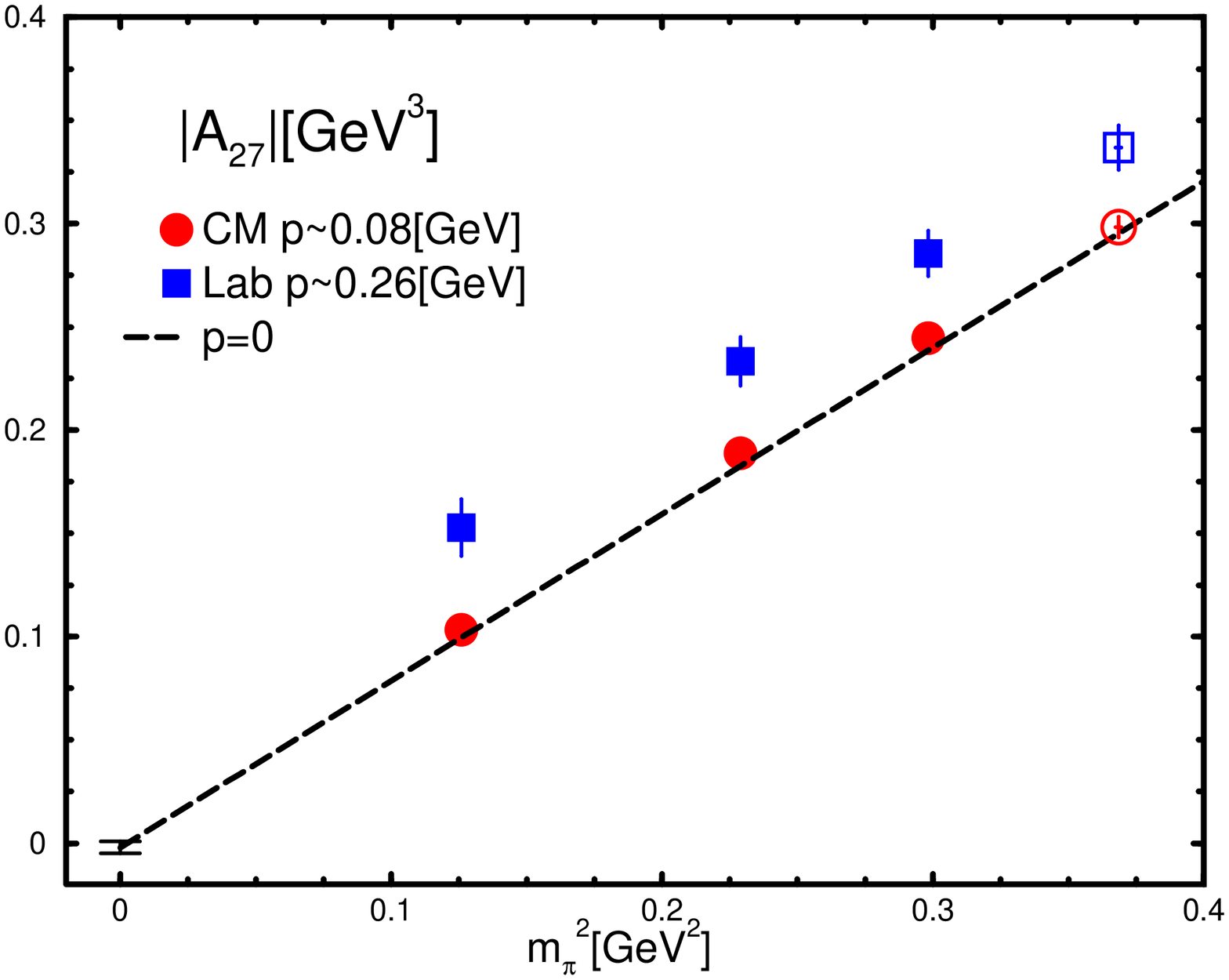}}}
\scalebox{0.28}[0.27]{
\rotatebox{270}{
\includegraphics{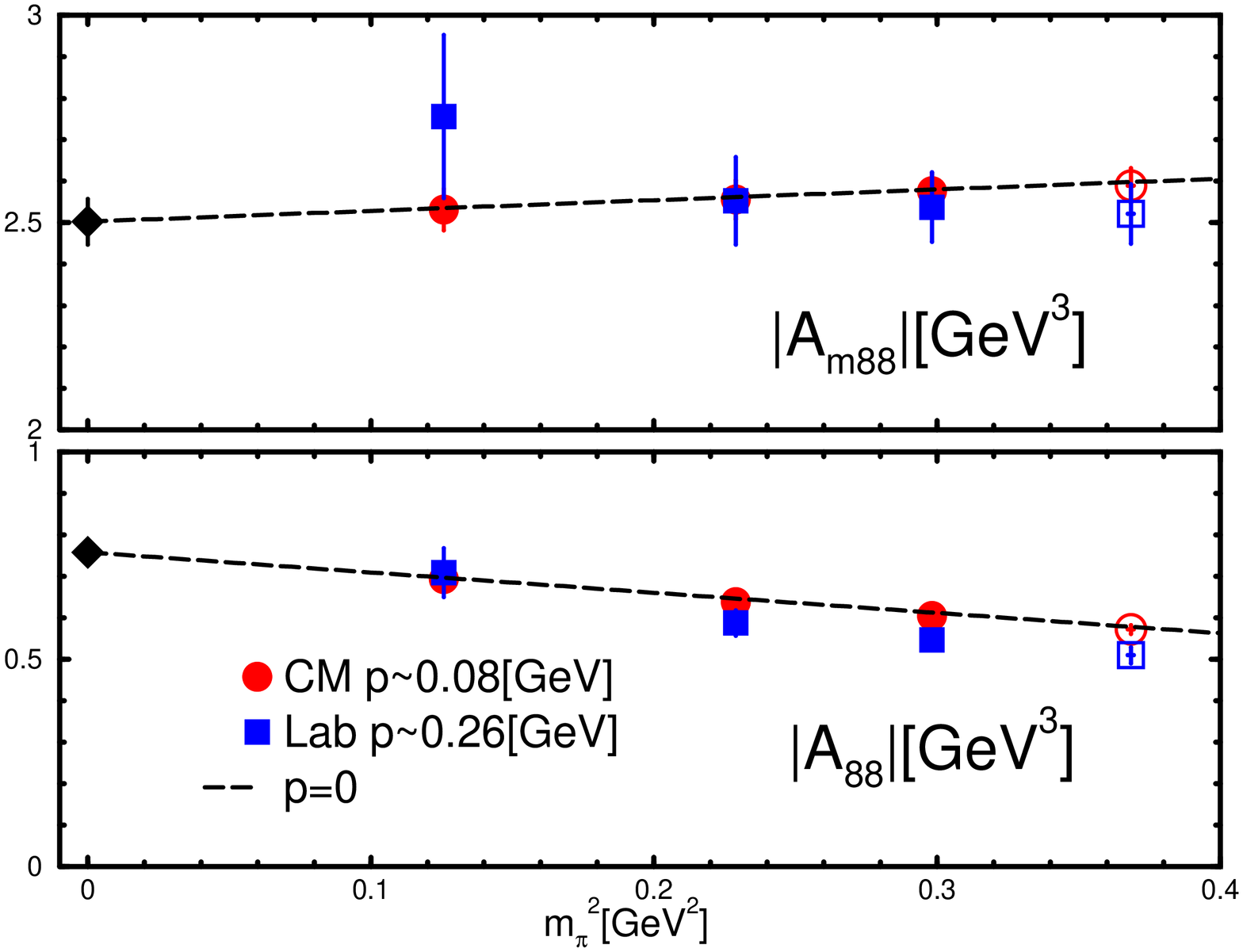}}}
\vspace{-5mm}
\end{center}
\caption{Infinite volume decay amplitudes 
obtained at different relative momenta.
Diamond symbols are fit results with $m_\pi^2 = p^2=0$.
Open symbols are omitted in global fit.
\label{fig_wme}}
\end{figure}
\begin{figure}[!h]
\begin{center}
\scalebox{0.32}[0.32]{
\rotatebox{270}{
\includegraphics{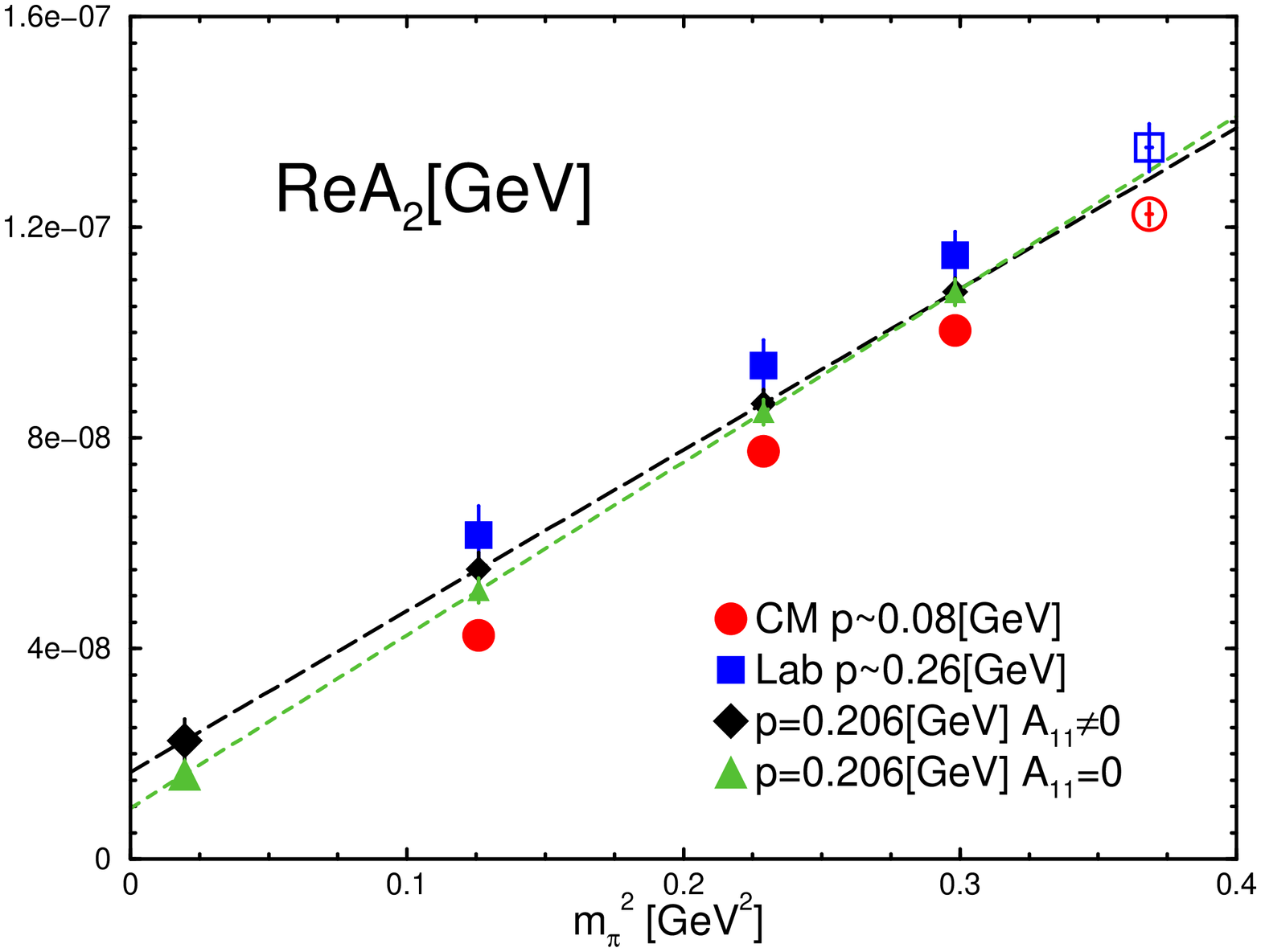}}}
\scalebox{0.32}[0.32]{
\rotatebox{270}{
\includegraphics{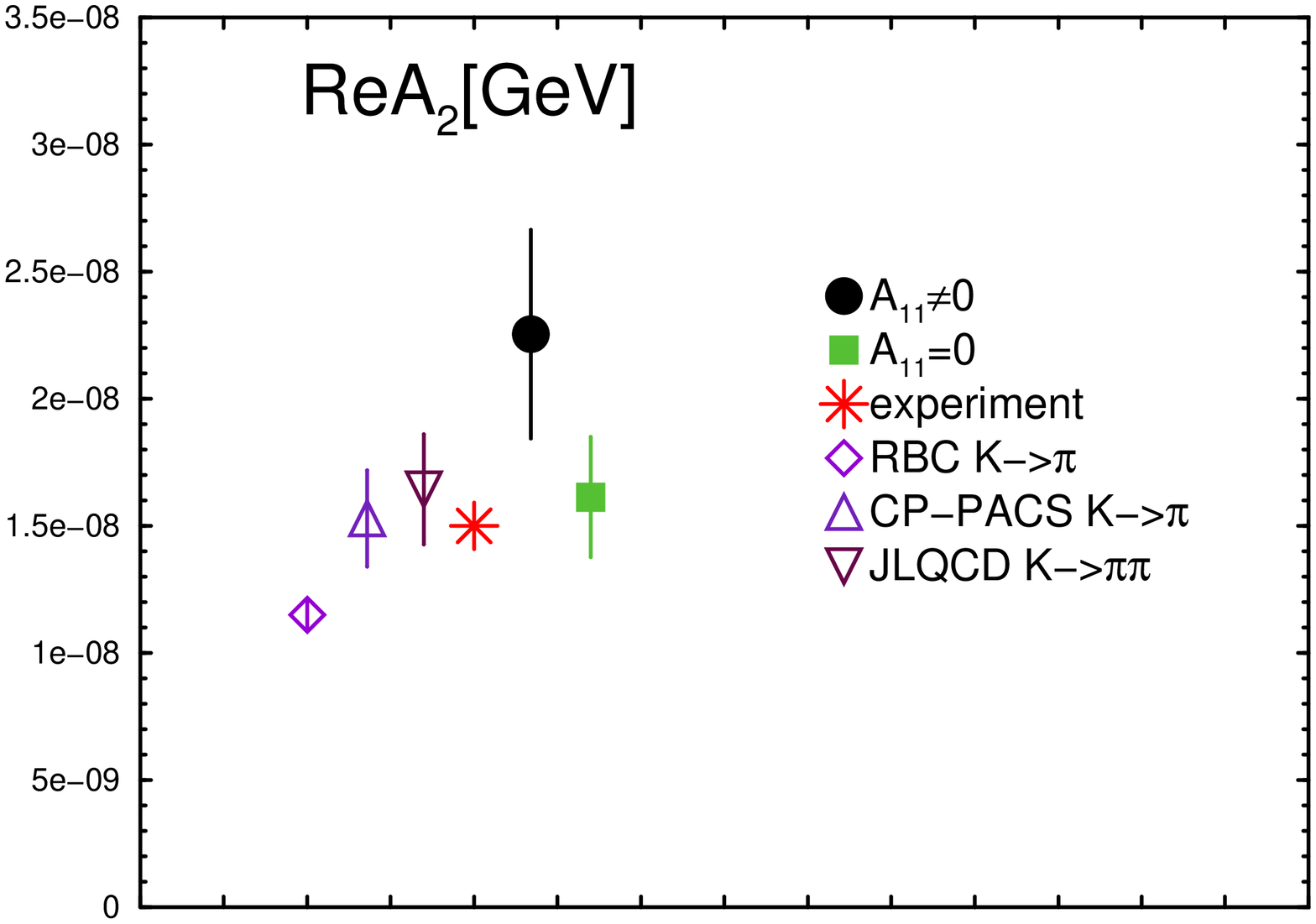}}}
\vspace{-5mm}
\end{center}
\caption{Measured $\mathrm{Re}A_2$ and result from previous works.
Open symbols are omitted in global fit.
In left panel dashed lines denote fit results.
\label{fig_ReA2}}
\end{figure}

The infinite volume decay amplitudes of all the operators 
are plotted in Fig.~\ref{fig_wme}.
The result of the 27 operator seems to vanish at the chiral limit
with zero relative momentum,
while the other elements remain a constant in the limit.
These trends of the pion mass dependence
are reasonably consistent with the prediction of 
ChPT at leading order~\cite{BGLLMPS}.
In order to investigate the $m_\pi^2$ and relative momentum $p^2$ dependences,
we carry out a global fitting for each decay amplitude
for $m_\pi^2$ and $p^2$ assuming a simple polynomial form as
\begin{equation}
A_{00} + A_{10} m_\pi^2 + A_{01} p^2 + A_{11} m_\pi^2 p^2.
\label{fit_func}
\end{equation}
The $\chi^2$/d.o.f. of the 27 operator is larger than 5
with all the four pion mass data,
so that we exclude the heaviest pion mass data in all the operators 
from the following analysis
to make the $\chi^2$/d.o.f. reasonable.
The constant term of the 27 operator is consistent with zero 
within the error, $A_{00}^{27} = -0.0020(29)$, as expected.
In other operators, we employ the same fitting function as
the 27 operator except $A_{11} = 0$.
The fit results are plotted in the figures,
and those at the both limits, 
$m_\pi^2 = p^2 = 0$, 
are represented by diamond symbols.
It should be noted that while the linear fit to $|A_{27}|$ is the
correct leading CHPT behavior, it is being used in a region of quite
large mass.  
Likewise the linear fit to $|A_{88}|$ and $|A_{m88}|$
omits possible logarithm terms which are of the same order.

The weak matrix elements are determined using
non-perturbative matching factors previously calculated
with the regularization independent (RI) scheme in Ref.~\cite{CK}.
The electroweak contributions of $\varepsilon^\prime / \varepsilon$,
$\langle Q_7 \rangle_2$ and $\langle Q_8 \rangle_2$,
are evaluated from $|A_{88}|$ and $|A_{m88}|$.
At the physical pion mass $m_\pi = 0.14$ GeV and relative momentum 
$p = 0.206$ GeV,
we obtain the matrix elements $\langle Q_7 \rangle_2 (\mu) = 0.2473(64)$ 
and $\langle Q_8 \rangle_2 (\mu) = 1.160(31)$ GeV$^3$
at the scale $\mu = 1.44$ GeV using the same fitting form as 
in the $|A_{88}|$ and $|A_{m88}|$ cases.

We calculate $\mathrm{Re}A_2$ from the weak matrix elements in the RI scheme
with
the Wilson coefficients evaluated by NDR scheme calculated in Ref.~\cite{CK}.
The left figure in Fig.~\ref{fig_ReA2} shows $\mathrm{Re}A_2$ obtained 
from the CM and Lab frame calculations.
The measured values of $\mathrm{Re}A_2$ are almost same 
as those of the 27 decay amplitude, in Fig.~\ref{fig_wme},
apart from the overall constant.
The reason is that the main contribution of $\mathrm{Re}A_2$
comes from the 27 amplitude.
We carry out a global fit using 
the same polynomial assumption~eq.(\ref{fit_func}) 
except the constant term $A_{00} = 0$
to evaluate the result at the physical pion mass $m_\pi = 0.14$ GeV
and momentum $p=0.206$ GeV.
In the fitting we omit the data at the heaviest pion mass
because of the same reason mentioned in the above.
We can also carry out a reasonable fitting, $\chi^2/$d.o.f. $=$ 1.2,
without $A_{11}$.
The fit results at each pion mass and the physical point
are plotted in the figure.

We plot the results of Re$A_2$ at the physical point
in the right panel of Fig.~\ref{fig_ReA2}
together with those of the previous works using
the indirect method~\cite{RBC,CP-PACS} and
direct calculation with evaluating the finite volume, 
two-pion interaction effect through ChPT~\cite{JLQCD}.
We estimate $\mathrm{Re}A_2 = 2.26(41)$ and 1.61(24)$\times 10^{-8}$ GeV 
at the physical point with and without $A_{11}$
term, respectively.
These results agree with each other and the experiment 
within two standard deviations.
The result is encouraging, albeit it includes systematic errors 
due to quenched approximation,
finite lattice spacing effects, and heavy pion and kaon masses.

\section*{Acknowledgments}

We thank Changhoan Kim for his previous study upon which 
the present work is based, and also thank RIKEN BNL Research Center, 
BNL and the U.S. DOE for providing the facilities essential for 
the completion of this work.

\providecommand{\href}[2]{#2}\begingroup\raggedright

\end{document}